\documentclass[twocolumn]{aastex701}
\submitjournal{ApJ}

\begin{document}

\title{Learning Galaxy Kinematics from Spatially Unresolved Spectra}

\author[orcid=0000-0002-8613-3140,sname='Waldrop']{Preston G. Waldrop}
\affiliation{School of Physics, Georgia Institute of Technology, 837 State St NW, Atlanta, GA 30332, USA}
\email[show]{pwaldrop3@gatech.edu}

\author[orcid=0000-0003-1035-3240,sname='Psaltis']{Dimitrios Psaltis}
\affiliation{School of Physics, Georgia Institute of Technology, 837 State St NW, Atlanta, GA 30332, USA}
\email{dpsaltis3@gatech.edu}

\author[orcid=0000-0003-4413-1523,sname='\"Ozel']{Feryal \"Ozel}
\affiliation{School of Physics, Georgia Institute of Technology, 837 State St NW, Atlanta, GA 30332, USA}
\email{fozel3@gatech.edu}

\author[orcid=0000-0002-3531-9842,sname='G\"uver']{Tolga G\"uver}
\affiliation{Istanbul University, Science Faculty, Department of Astronomy and Space Sciences, Beyaz\i t, 34119, \.Istanbul, T\"urkiye}
\affiliation{Istanbul University Observatory Research and Application Center, Istanbul University 34119, \.Istanbul, T\"urkiye}
\affiliation{School of Physics, Georgia Institute of Technology, 837 State St NW, Atlanta, GA 30332, USA}
\email{tolga.guver@istanbul.edu.tr}

\begin{abstract}
Galaxy kinematics encode information about the assembly histories of galaxies and the black holes at their centers, motivating the measurement of kinematic parameters for large galaxy populations. In particular, such parameters have been shown to help identify galaxies that host supermassive black hole binaries, a major outstanding question in multi-messenger astrophysics. Yet, inferring kinematic properties from spatially unresolved observations, where distant galaxies appear as point sources, remains a challenge. Spectroscopy can address this challenge because Doppler effects encode internal motions into spectral line profiles. Exploiting this encoding requires knowing how spatial structures map onto the line profiles but this cannot be derived from first principles for complex systems such as galaxies. We present a neural network trained on spatially resolved spectra that infers kinematic properties from spatially integrated spectra, offering two advantages over classical approaches: it learns the kinematic encoding directly from data without requiring spatial resolution and remains effective even when spectral resolution only marginally resolves the lines. This enables recovering the kinematic properties for millions of unresolved galaxies in existing and future surveys.
\end{abstract}

\keywords{\uat{Galaxies}{573} --- \uat{Galaxy mergers}{608} --- \uat{Galaxy Kinematics}{602} --- \uat{Supermassive black holes}{1663} --- \uat{Neural Networks}{1933} --- \uat{Algorithms}{1883}}


\section{Introduction} \label{sec:intro}

Contrary to nearby systems that can be studied through spatially resolved methods, the vast majority of astronomical observations reduce complex, extended structures to point sources, collapsing spatial information into a limited set of integrated measurements. This creates a central inference problem: how to recover the underlying multi-dimensional structure, dynamics, and kinematics of a system from observations that lack direct spatial resolution. Historically, spectroscopy has provided one of the few avenues to address this challenge, using Doppler-induced distortions of spectral lines to encode information about internal motions and spatial structure.

A widely used example of inferring internal structure and kinematics from unresolved observations is the analysis of stellar spectral line profiles~\citep{Mihalasbk}. Stellar rotation imparts a characteristic broadening to spectral lines that can be modeled to infer the projected rotational velocity~\citep[see, e.g.,][]{Shajn1929,SimonDiaz2007,Fremat2023}. Inhomogeneities on the stellar surface, such as starspots, produce localized dips in line profiles that encode their position and contrast~\citep[see, e.g.,][]{Vogt1983,Strassmeier2009}. Stellar winds produce the characteristic P-Cygni line profile, where a redshifted emission component and blueshifted absorption component together encode the radial velocity structure of the outflow~\citep[see, e.g.,][]{Castor1979,Kudritzki2000,Hawcroft2024}. In each case, knowledge of the physical process allows for the spatial structure to be forward-modeled and compared to the observed line profile.

Exploiting spectral line profiles to infer internal structures and kinematics requires prior knowledge of the mapping between the physical properties of the source and the resulting line profile. For the stellar cases described above, this mapping is relatively tractable due to the spherical symmetry of stars and the geometrically simple nature of rotation and outflows. Such first-principles approaches become much harder to follow for more complex systems. For galaxies, the vast majority of sources in current spectroscopic surveys are spatially unresolved. However, using such observations to infer kinematic properties is challenging both because galaxies are three-dimensional, differentially rotating systems and because they have more complex and diverse internal structures than stars.

Dynamics and kinematics of galaxies contain information about their assembly histories and central black holes. Of particular interest to the present study, the kinematic parameters of galaxies have been shown to help identify galaxies that host the supermassive black-hole binaries, such as those we detect with gravitational-wave observatories, a major outstanding question~\citep{Tamara2022, Bardati2024b}. In particular, large-scale galaxy simulations have demonstrated a correlation between galaxies that host supermassive black-hole binaries and the magnitude of the spin parameter $\lambda_R$, which is a proxy for the projected angular momentum of the galaxy~\citep{Bardati2024b, emsellem2007, Romulus25}. Identifying such hosts throughout cosmic history could, therefore, be achieved by measuring the spin parameter across large populations of galaxies spanning a wide range of distances from Earth.

In principle, it is possible to measure the spin parameter $\lambda_R$ directly using  Integral Field Unit Spectrographs (IFUs)~\citep{bacon2001}. These instruments allow a measurement of the velocities of gas and stars across the face of each galaxy, which can then produce kinematic maps~\citep{manga}. Integrating this map spatially quantifies the spin parameter of the galaxy~\citep{emsellem2007}. The requirement for spatially resolved spectra, however, limits the method to galaxies that are nearby (i.e., with redshifts $z\leq 0.1$) so that they can be resolved with existing bundles of optical fibers placed in the focal planes of telescopes. 

This seemingly technical limitation has significant consequences. The galaxies most likely to harbor supermassive black hole binaries lie well beyond this redshift limit~\citep{Krause2025}. Even within the local universe, IFU surveys are time consuming and observationally expensive, resulting in small existing datasets, with as few as 10,000 galaxies~\citep{manga}. By contrast, spatially unresolved spectroscopic surveys, such as eBOSS~\citep{eBOSS} and DESI~\citep{DESI2026A}, contain millions of galaxies across a much wider range of cosmic distances. In principle, they could enable kinematic characterization at an entirely different scale, if the spin parameter could be extracted from spatially integrated spectra.

In this work, we develop a machine learning framework to infer the spin parameter $\lambda_R$ directly from spatially unresolved galaxy spectra. Rather than constructing an explicit forward model for the mapping between velocity fields and spectral line profiles, which becomes intractable for realistic galaxy structures, we take a data-driven approach. We use the fact that, although their detailed properties are complex, galaxies are not arbitrary systems; they exhibit regular behavior, and simple empirical descriptions of their surface brightness and velocity profiles have been established over decades of observations~\citep{Sofue2001,Graham2005}. These regularities leave systematic imprints on integrated spectra, making them a natural target for neural network architectures.

\begin{figure*}[t]
  \centering
  \includegraphics[width=1.0\textwidth]{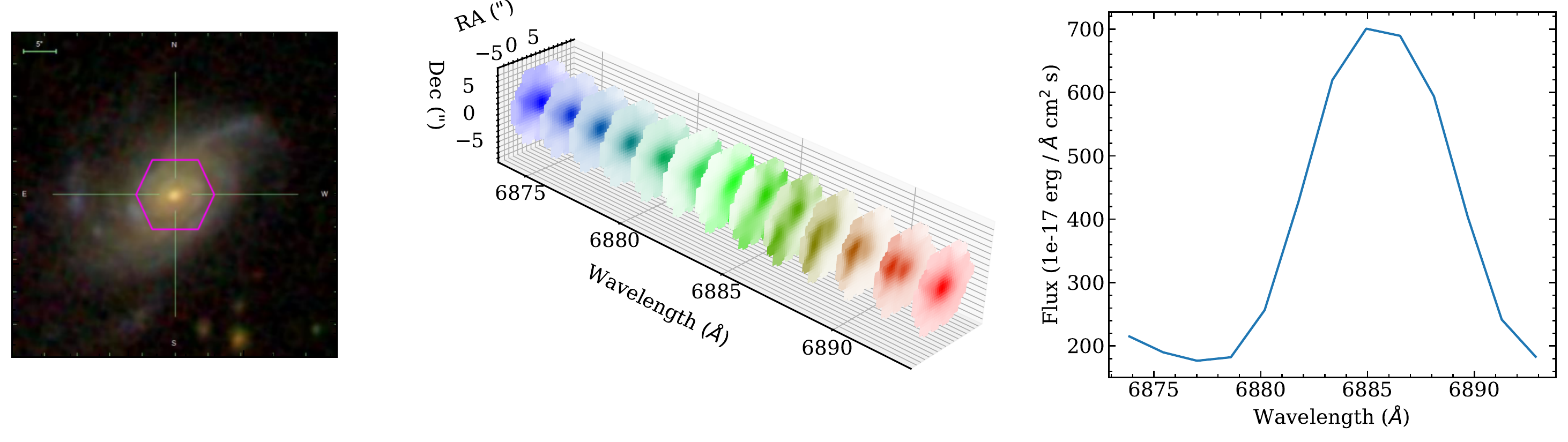}
  \caption{The data preparation used to train a neural network to learn kinematic information from spatially unresolved spectroscopic data. {\em (Left)\/} Image of \texttt{MaNGA} target galaxy corresponding to PlateIFU 10213-1902. {\em (Middle)\/} Three-dimensional representation of the flux data cube, limited to within 10~\AA\ of the Hydrogen alpha line. Different colors correspond to the brightness of the image of the galaxy at different wavelengths. {\em (Right)\/} Spatially integrated flux as a function of wavelength. }
  \label{fig:3d2d}
\end{figure*} 

In our work, we train a fully connected network on galaxies with IFU spectroscopy in the \texttt{MaNGA} survey~\citep{manga} to learn the mapping between spatially resolved kinematics and spatially integrated spectra, and assess its ability to recover $\lambda_R$ from unresolved observations. We compute $\lambda_R$ from the H$\alpha$ gas kinematics, since H$\alpha$ is a prominent emission line whose profile can be measured without a detailed fit to the stellar continuum. We find that the network successfully infers kinematic properties directly from integrated spectra and remains effective even when the spectral instruments of a large survey only marginally resolve the line profiles. 

\begin{figure*}[t]
  \centering
  \includegraphics[width=1.0\textwidth]{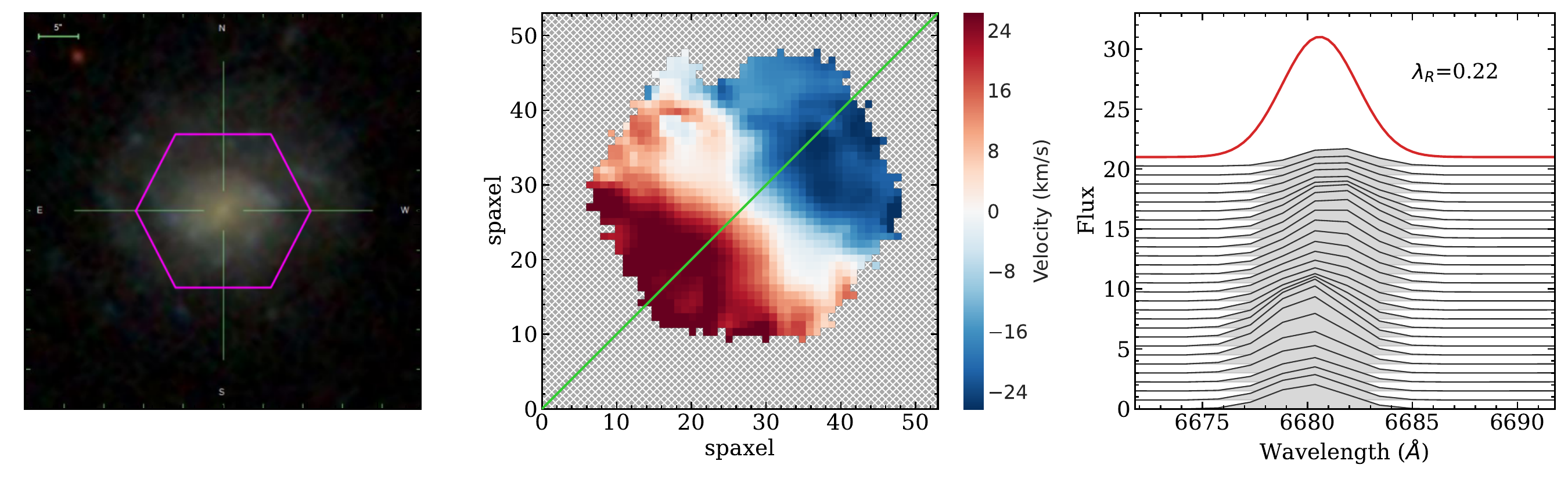}
  \includegraphics[width=1.0\textwidth]{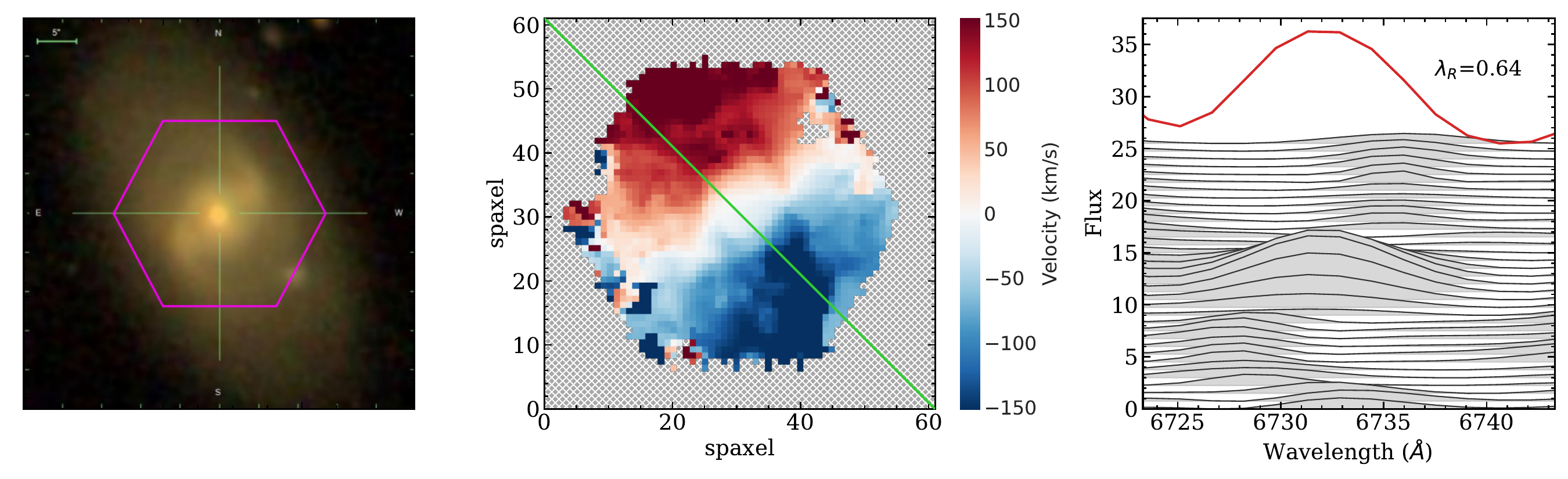}
  \caption{(Left) Images of two \texttt{MaNGA} target with IFU overlay. (Middle) Maps of line-of-sight velocities. (Right) Grey lines show the scaled and offset H$\alpha$ line profiles from the spaxles along the green cross sections shown in the middle panels. The red lines are the scaled, spatially integrated H$\alpha$ line profiles across all spaxles in each image. The top row corresponds to the low-spin galaxy with PlateIFU 10214-6103 and $\lambda_R=0.22$; the bottom row corresponds to the high-spin galaxy with PlateIFU 12088-9102 and $\lambda_R=0.64$. Larger values of the spin parameter $\lambda_R$ correspond to a broader range of line-of-sight velocities, which broaden and alter the shapes of the spatially integrated line profiles.}
  \label{fig:lambda}
\end{figure*}

\section{Measuring Spin Parameters with IFUs} \label{sec:pw}

IFUs use bundles of optical fibers to obtain spectra across different regions of a galaxy, producing spatially resolved spectral elements (spaxels). These measurements form data cubes of shape $(M, N, N)$, where $M$ is the number of spectral bins in wavelength and $N \times N$ samples the two-dimensional spatial field~\citep{Law2016,Yan2016}. Each resulting data cube encodes the flux as a function of wavelength and sky position (right ascension and declination), enabling a direct reconstruction of the kinematic structure of the galaxy. Existing IFU surveys include the SAMI Galaxy Survey with $\sim 3000$ galaxies~\citep{SAMI2015} and \texttt{MaNGA} with $\sim 10000$ galaxies~\citep{manga}.

IFU surveys are restricted to nearby galaxies because spatially resolved spectroscopy requires the targets to subtend a sufficiently large angular size on the sky. This limitation follows from the geometric relation  $\theta \approx 2R/D$, where $\theta$ is the angular aperture of the instrument, $R$ is the radius of the galaxy, and $D$ is the distance from Earth. For the effective radius of Milky Way-like galaxy, $\sim$5~kpc, this limit becomes
\begin{equation}
    D \approx 150 \left( \frac{R}{5~\text{kpc}} \right) \left( \frac{\theta}{14"} \right)^{-1}~\text{Mpc}\;.
\end{equation}

Using such IFU spectra, we can measure the spin parameter, which quantifies the degree to which a galaxy is supported by rotation (high $\lambda_{R})$ versus the random motion of the gas and stars (low $\lambda_{R}$). It was introduced by \citet{emsellem2007} as a proxy for the projected angular momentum of the galaxy and is defined as
\begin{equation}
    \lambda_R \equiv \frac{\langle R |V| \rangle}{\langle R \sqrt{V^2 + \sigma^2} \rangle},
    \label{eq:emslam}
\end{equation}
where the numerator and denominator are luminosity weighted values across the face of the galaxy, $R$ is the projected distance from the center, $V$ is the line-of-sight velocity, and $\sigma$ is the velocity dispersion. \citet{Bardati2024b} showed that low values of $\lambda_R$ are indicative of recent major galactic mergers, reflecting the conversion of ordered rotational motion into random motions during these events.

\begin{figure}[tp]
  \centering
\includegraphics[width=0.48\textwidth]{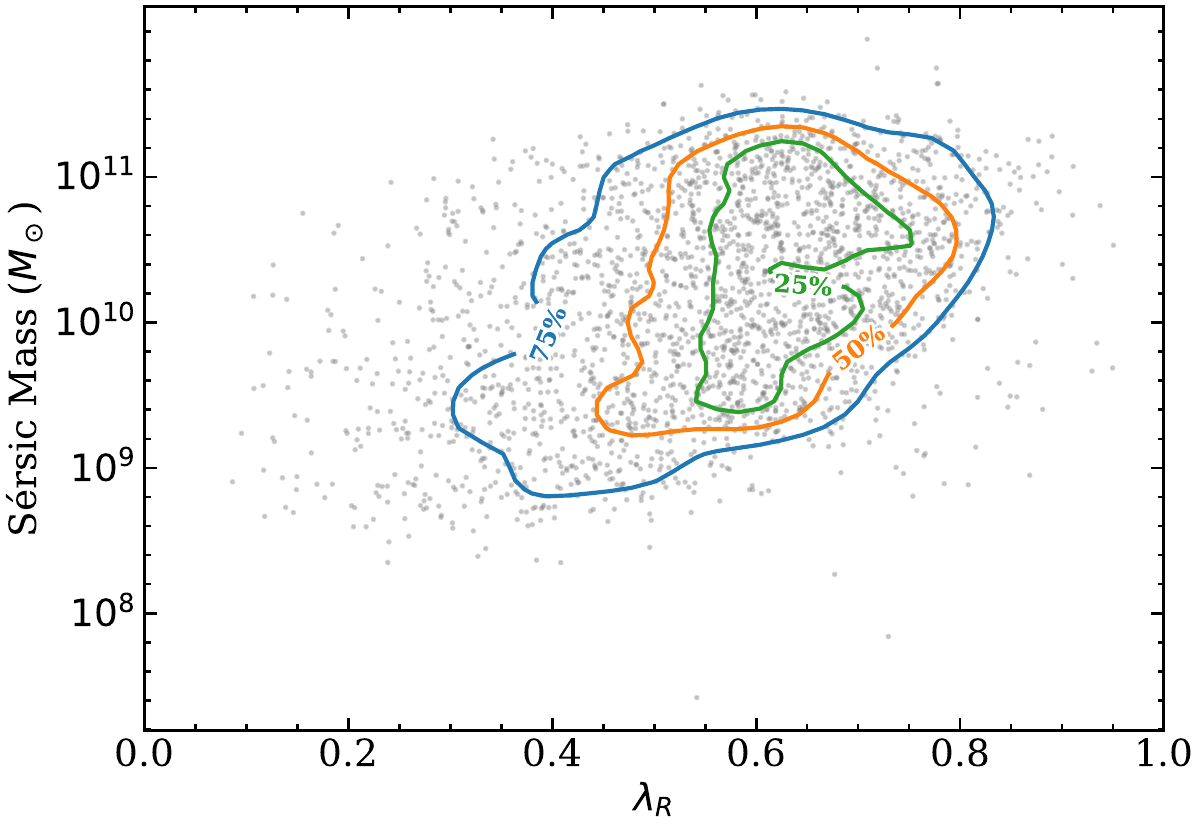}\vfill 
\includegraphics[width=0.48\textwidth]{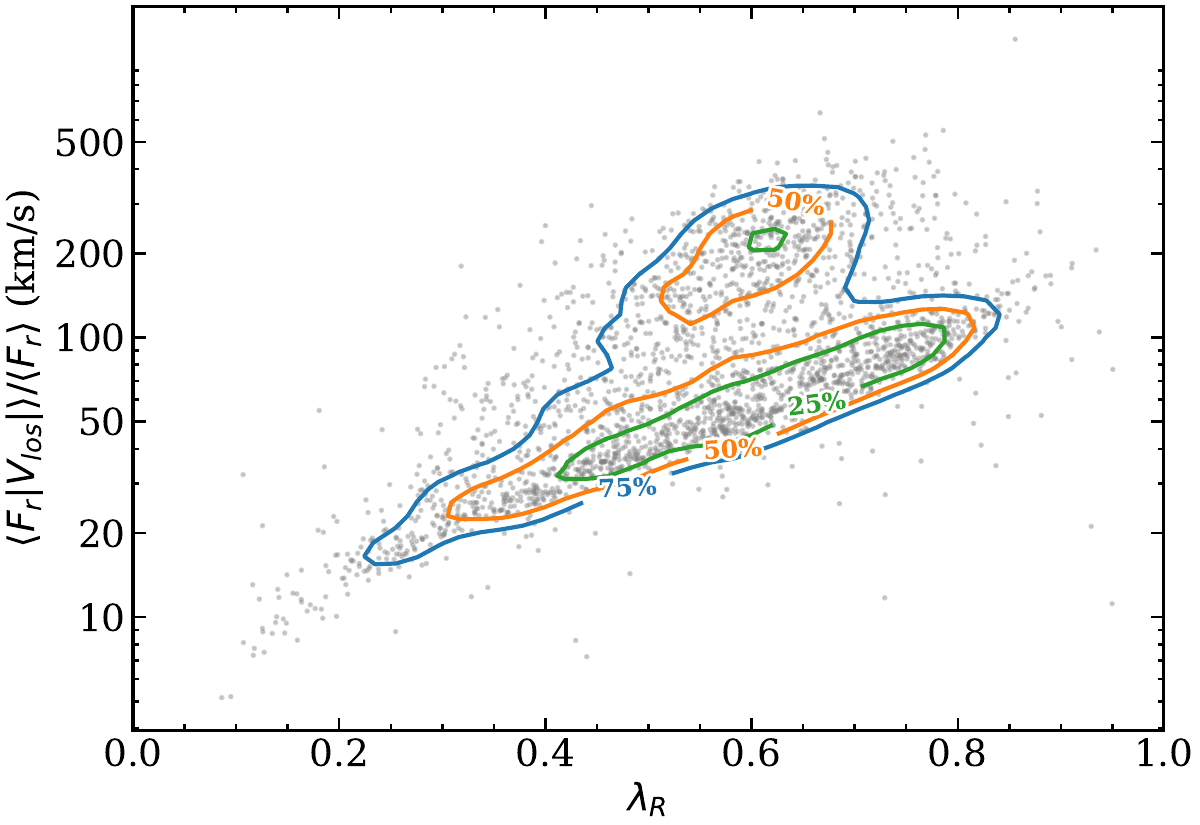} \vfill
\includegraphics[width=0.48\textwidth]{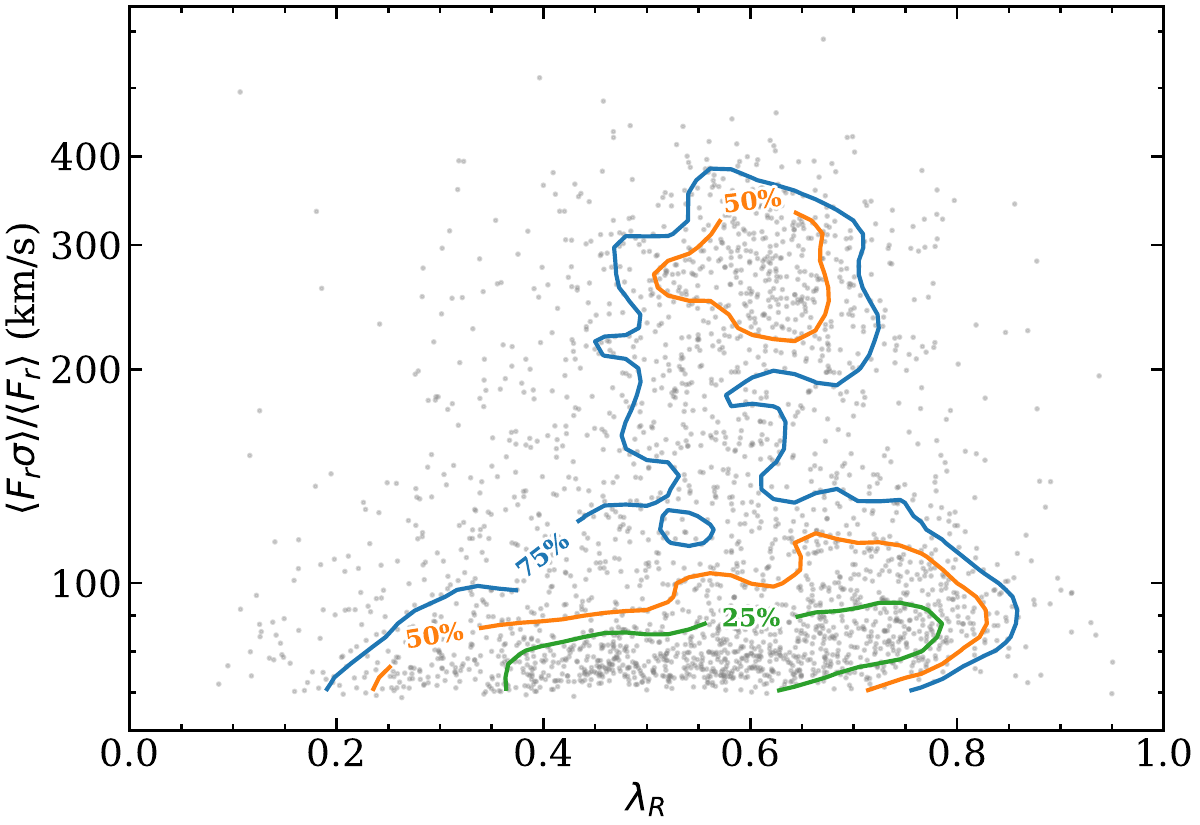}
\caption{{\em (Top)\/} The S\'ersic mass, {\em (Middle)\/} the r-band flux weighted average of line-of-sight velocity, and {\em (Bottom)\/} the r-band flux weighted average velocity dispersion of each galaxy in the training sample plotted against their spin parameter $\lambda_R$. In all three panels, the green, orange, and blue contours correspond to the 25-th, 50-th, and 75-th percentiles of galaxies, respectively. The lack of tight correlation between these variables indicates that no single global kinematic measure or galaxy property adequately predicts the value of the spin parameter $\lambda_{\rm R}$.} 
  \label{fig:confusion}
\end{figure}

\section{Data Selection and Preparation} \label{sec:data}

In this study, we use data from the Mapping Nearby Galaxies at Apache Point Observatory (\texttt{MaNGA}) integral field unit (IFU) survey \citep{manga}, conducted as part of the Sloan Digital Sky Survey IV (SDSS-IV) \citep{sdss4}. \texttt{MaNGA} observations were obtained using the 2.5-meter Sloan Foundation Telescope and the BOSS spectrographs \citep{sdssTeles, eBOSSspectro, mangaFeed}. The data are publicly available in SDSS Data Release 17 \citep{dr17}. The survey includes approximately ten thousand galaxies in the nearby universe, covering 2700~deg$^2$ of the sky, out to a redshift of $z\sim 0.03$. This redshift limit corresponds to a distance of $\sim 130$~Mpc for standard $\Lambda$CDM cosmology. 

The IFU observations provide a spectrum at $\sim\nobreak 720–3500$ spatial pixels (“spaxels”) across the face of each galaxy, forming a three-dimensional data cube that can also be viewed as a series of galaxy images at different wavelengths (see Fig.~\ref{fig:3d2d}). For each spaxel, we use the line-of-sight velocities and velocity dispersions provided by \texttt{MANGA}~\citep{mangaDAP}. Using these measurements from the spatially resolved spectra, we calculated the spin parameter for each galaxy using the discrete version of Equation~(\ref{eq:emslam}), i.e.,
\begin{equation}
    \lambda_R = \frac{\displaystyle{\sum_{n}} F_n r_n |v_n|}{\displaystyle{\sum_{n}} F_n r_n \sqrt{v_n^2 + \sigma_n^2}},
    \label{eq:lambda}
\end{equation}
where $F_n$ is the r-band flux, $r_n$ is the projected radius on the face of the galaxy, $v_n$ is the line-of-sight velocity, and $\sigma_n$ is the velocity dispersion of the $n$-th spaxel. To perform this calculation, we employ existing data cubes and kinematic maps using the Python package \texttt{Marvin} provided by SDSS \citep{marvin}.

We then generate spatially unresolved spectra by integrating the spaxels along RA and DEC. The dataset for each galaxy is, therefore, reduced from an array defined by $(M,N,N)$ to a single vector of size $M$, where $M$ is the number wavelength bins and $N$ is the number of spatial pixels along each direction. We note that the generated spectra match the spectral resolution of the eBOSS survey as that survey uses the same spectrograph as \texttt{MaNGA}. This will allow us to apply our approach directly to the galaxies contained in the eBOSS survey~\cite{eBOSS}; in \S5, we also discuss applications to other surveys such as \texttt{DESI}~\cite{DESI}.

We then selected galaxies with {\it (i)} no data quality flags and {\it (ii)} a well-defined H$\alpha$ emission line. We chose the H$\alpha$ line for this initial proof-of-principle study because it is ubiquitous and typically strong, allowing its kinematic signatures to be measured without detailed spectral modeling. In future work, this analysis can be extended to other gas emission lines and stellar absorption features. 

Our criteria yielded 1471 galaxies out of the $\sim$10,000 in the \texttt{MaNGA} survey. The resulting sample spans a broad range of spin parameters, $0.1 \le \lambda_R \le 0.9$, ensuring representation of both dispersion-dominated and rotation-dominated systems. Finally, we divided this sample randomly into a training set of 1176 galaxies and a test set of 295 galaxies. The distributions of $\lambda_R$ in the training and test sets are comparable. 

We examined whether our sample exhibits correlations between $\lambda_R$ and other galaxy properties that could introduce biases into or confound the neural network training. In particular, previous studies have found that more massive galaxies tend to have lower $\lambda_R$, reflecting, in part, their merger-driven assembly histories \citep{emsellem2007, emsellem2011}. However, as we show in the top panel of Figure~\ref{fig:confusion}, our training sample exhibits a broad distribution in both the S\'ersic-derived stellar mass and $\lambda_R$, without a strong correlation between the two. This reduces the risk that the neural network could exploit mass-dependent trends instead of learning the kinematic information encoded in the spectra. We also examined the relationship between $\lambda_R$ and two simpler kinematic measures: the $r$-band flux-weighted line-of-sight velocity (middle panel) and the average velocity dispersion (bottom panel). The absence of strong correlations suggests that no single global kinematic measure adequately predicts $\lambda_{\rm R}$, further supporting the need to extract more detailed kinematic information from the spectra.

\begin{figure}[t]
  \centering  
 \includegraphics[width=0.48\textwidth]{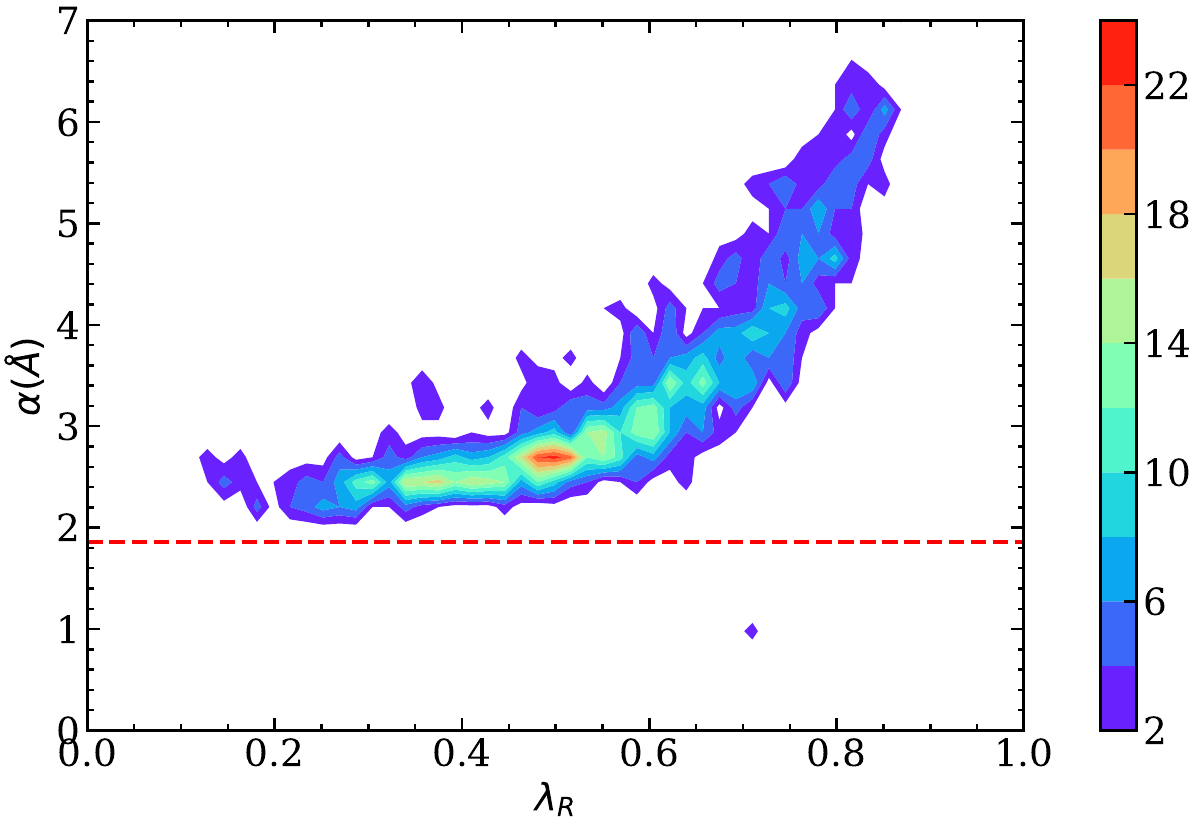}
    \caption{Contours of the H$\alpha$ line width, $\alpha$, in the integrated spectrum, obtained from fits with a generalized Gaussian profile, as a function of the spin parameter $\lambda_R$ for 1176 galaxies. The color bar indicates the number of galaxies per bin. The red dashed line marks the spectral resolution of the SDSS BOSS spectrograph. A direct mapping between $\alpha$ and $\lambda_R$ is limited by instrumental resolution, which suppresses sensitivity to values of $\lambda_R\le 0.7$.}
  \label{fig:alpha}
\end{figure}

\subsection{Kinematic Imprints on Integrated Spectra} \label{subsec:lines}

The goal of this work is to explore whether spatially integrated spectral line profiles contain imprints of the underlying galaxy kinematics and, in particular, of the spin parameter $\lambda_R$, and to exploit any such imprints to learn galaxy kinematics. The impact on the line profiles is expected because, although spatial information is lost, Doppler shifts across different spaxels combine in a systematic way, encoding the distribution of line-of-sight velocities into the shape and width of emission lines. 

Figure~\ref{fig:lambda} provides an illustrative example of this effect. The top row shows a galaxy with a low spin parameter ($\lambda_R=0.22$), while the bottom row shows one with a high spin parameter ($\lambda_R=0.64$). The leftmost panels show the field-of-view of the IFU superimposed on the images of the galaxies. The middle panels show the line-of-sight velocity maps derived from the H$\alpha$ line across the face of each galaxy. The rightmost panels show the spectra of the spaxels along a diagonal cross section (shown in green lines), and the corresponding integrated spectrum. 

The low-spin galaxy exhibits little variation in line-of-sight velocity across the field, leading to nearly aligned H$\alpha$ line centroids and a narrow feature in the integrated spectrum. In contrast, the high-spin galaxy shows significant velocity gradients, producing Doppler-shifted line centroids across spaxels that combine into a broader H$\alpha$ profile when integrated. This behavior is the reason why the value of the spin parameter $\lambda_R$ is encoded in the profiles of spectral lines, justifying a data-driven approach to inferring this quantity from unresolved spectra.

\begin{figure*}[t]
  \centering
 \includegraphics[width=0.48\textwidth]{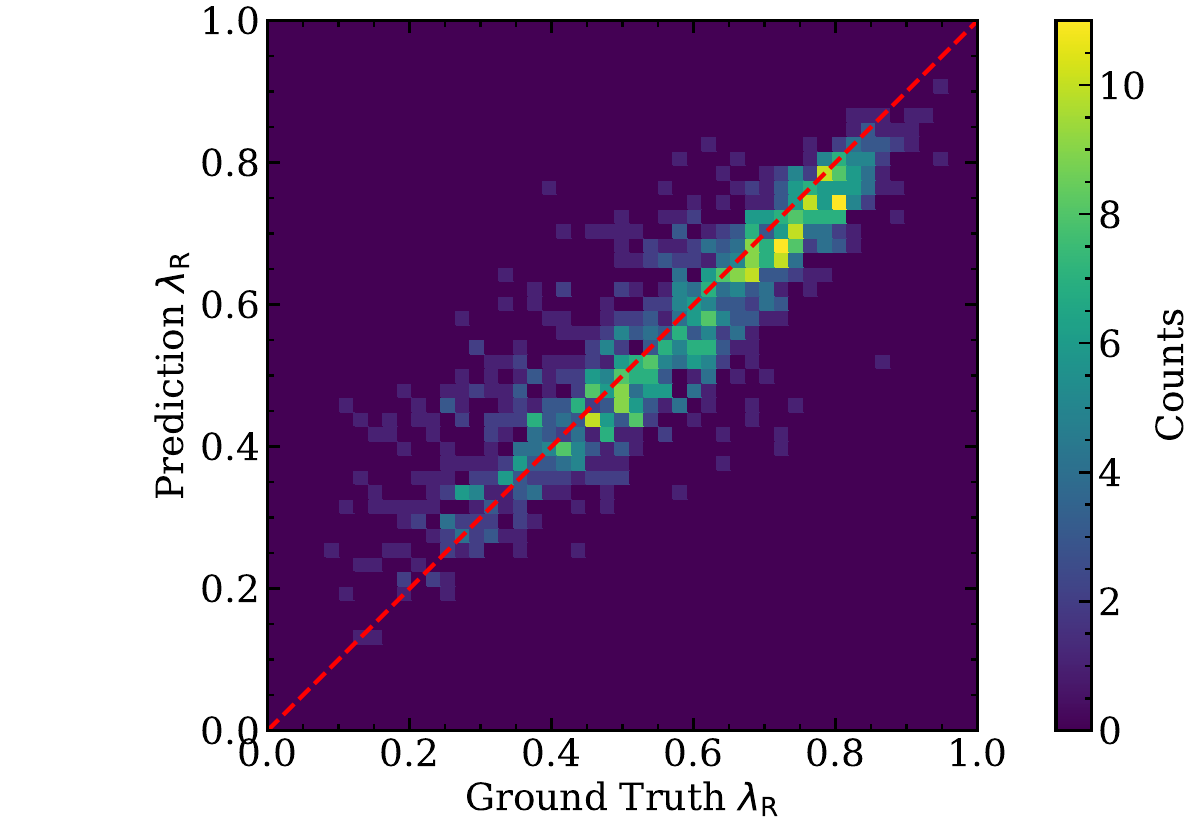}\hfill
 \includegraphics[width=0.48\textwidth]{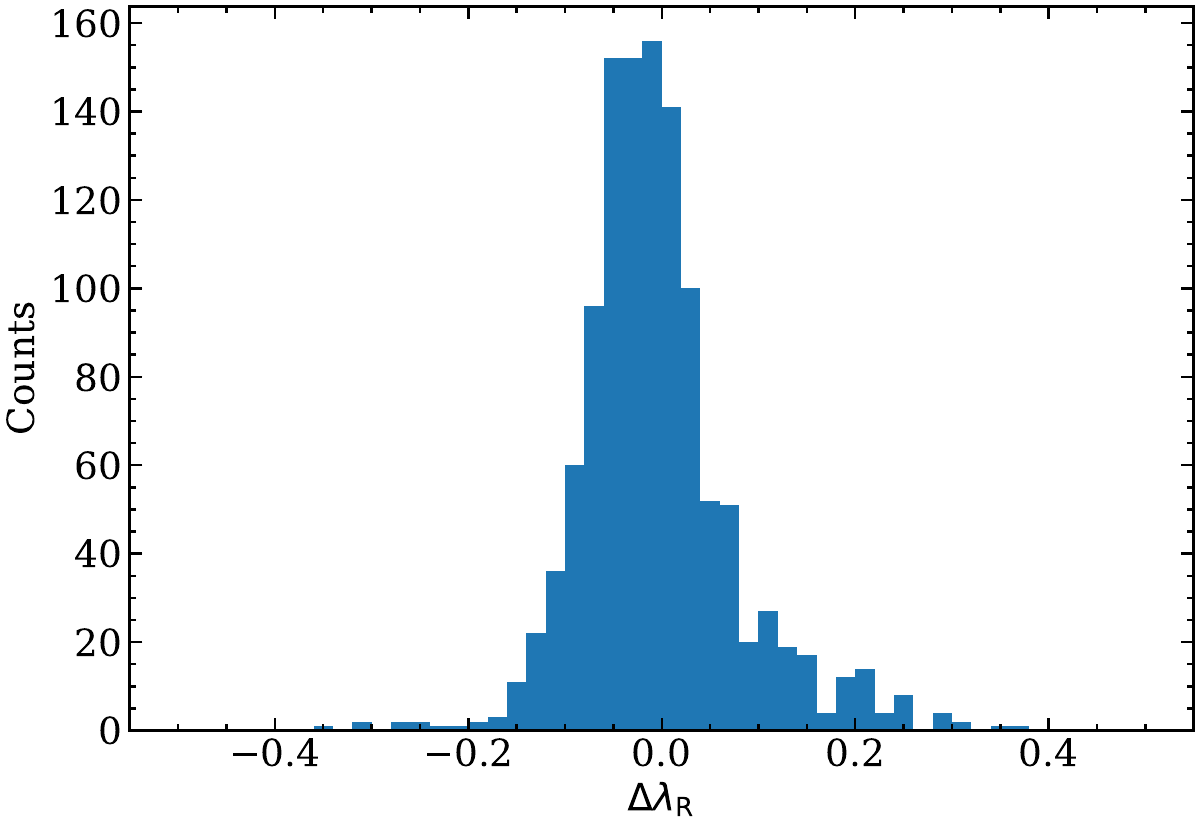}
  \caption{{\em (Left)\/} Two-dimensional histogram of the predicted $\lambda_R$ from the neural network versus ground truth $\lambda_R$. The color bar indicates counts. The red dashed line indicates the case when the prediction is equal to the ground truth. {\em (Right)\/} Histogram of the difference between the predicted and ground-truth values of the spin parameter $\lambda_R$ in our sample. The performance of the neural network is strong and does not degrade for small values of the spin parameter $\lambda_R$.}
  \label{fig:nn}
\end{figure*}

\subsection{The Need for a Machine-Learning Solution}

A traditional approach to extracting kinematic information from integrated spectra would be to adopt a parametric model for the line profile and to correlate its fitted parameters to the spin parameter $\lambda_R$. This would require a predictive model for the spatial distribution of line-of-sight velocities and surface brightness across galaxies, which together determine the emergent line profile as well as the value of the spin parameter (see Eq.[\ref{eq:lambda}]). Because such first-principles models are not available for realistic galaxy populations, we could instead use a flexible, phenomenological description of the line profile.

As an illustrative exercise, we fit the H$\alpha$ integrated line profile, $F(\lambda)$ with a generalized Gaussian distribution
\begin{equation}
F(\lambda) = I_c + I_0 \exp \left[ -\left( \frac{|\lambda - \mu|}{\alpha} \right)^\beta \right],
\end{equation}
where $\alpha$ measures the width of the line and $\beta$ its shape. Figure~\ref{fig:alpha} shows the relation between $\alpha$ and $\lambda_R$ we find for the galaxy sample introduced earlier. As expected, $\alpha$ increases with $\lambda_R$, reflecting the broadening of the line profile due to larger velocity gradients. However, this approach quickly encounters a limitation set by the spectral resolution of the instrument. Because the analytic profile is only an approximate description of the true line shape, the inference relies primarily on coarse features such as the overall width, which cannot be robustly constrained below the instrumental resolution. 

As seen in Figure~\ref{fig:alpha}, this leads to a saturation in the $\alpha$–$\lambda_R$ relation at values of $\lambda_R \lesssim 0.5$.   This limitation arises primarily from the inability of the simple analytic profiles to capture the structured line shapes produced by realistic galaxy kinematics as the intrinsic line widths approach the resolution limit.

\section{Neural Network Inference of the Spin Parameter} \label{sec:nn}

To quantify the extent to which the spin parameter $\lambda_R$ can be recovered from spatially integrated spectra, we train a neural network that learns the mapping between spectral features in the integrated line profile and the spin parameter. The input consist of a reduced representation of the spatially integrated H$\alpha$ line profile sampled across wavelength for each galaxy and the output is the corresponding value of $\lambda_R$ inferred from spatially resolved observations.

The architecture comprises 20 input nodes, two hidden layers with 128 neurons each using the sigmoid activation function, and a single output node. The network is trained to minimize the mean squared error between the predicted and true values of $\lambda_R$ using the Adam optimizer \citep{adam}. Training converges to a loss of $\sim 0.01$, with comparable performance on the test set, indicating no evidence for overfitting. For this network, the \texttt{MaNGA} observations were downloaded to a lab computer using $\sim 1$~Tb of storage and the training was also done locally.

Figure~\ref{fig:nn} shows a two-dimensional histogram of the predicted versus true $\lambda_R$ values. The network accurately recovers the full dynamic range of $\lambda_R$ and clearly separates low- and high-spin systems. In contrast to parametric approaches based on fitting analytic line profiles, the neural network does not exhibit a degradation in performance at low $\lambda_R$, where the intrinsic line widths approach the spectral resolution. This demonstrates that the network is able to exploit structure in the line profiles beyond simple width measures, effectively overcoming the limitations imposed by instrumental resolution. The rms difference between the predicted and ground-truth values of the spin parameter $\lambda_R$ is $\simeq 0.08$ (Fig.~\ref{fig:nn}).

A machine-learning model can learn the distribution of line profiles associated with different spin parameters and exploit subtle features beyond a single width parameter. This is analogous to measuring the centroid of a spectral line, where a known profile allows the line center to be determined with precision exceeding the nominal resolution of the instrument. By learning the appropriate basis of physically realized line profiles, a neural network can, therefore, extract kinematic information beyond the limits imposed by simple parametric fits.

\begin{figure}[t]
  \centering  
 \includegraphics[width=0.47\textwidth]{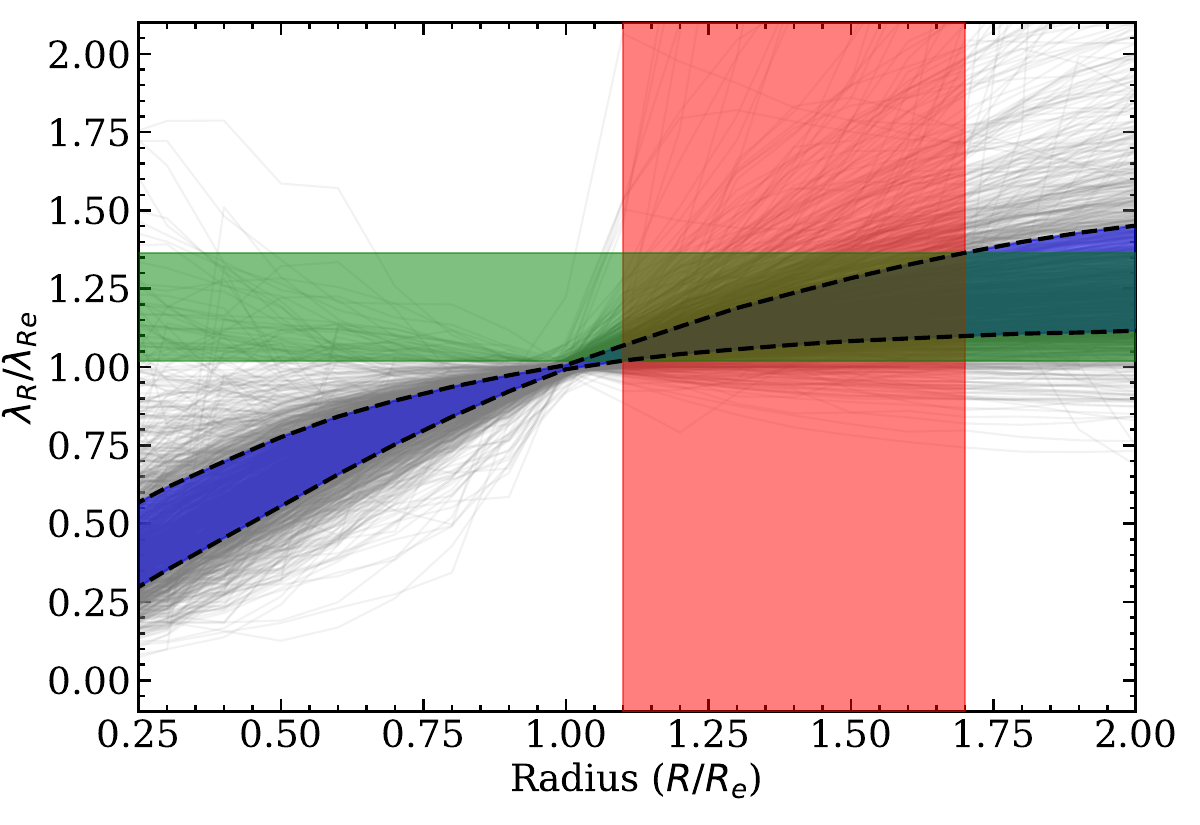}
    \caption{{\em (Grey lines)\/} Galaxy spin parameter as a function of radius of integration, for the training sample of galaxies. The radius is in units of the half-light radius, $R_e,$ and the spin parameter is normalized by the spin parameter at the half-light radius, $\lambda_{R_e}.$ The blue region outlines the 20th to 80th percentiles of the spin parameter, while the red region corresponds to radii from $1.1~R_e$ to $1.7~R_e$. The green region shows that there is a $\pm$20\% uncertainty on the measure of $\lambda_R$ and a $+20\%$ bias of $\lambda_R$ measured at $>1.7~R_e$ as apposed to $\lambda_{R_e}$.}
  \label{fig:lamR}
\end{figure}

\section{Discussion} \label{sec:concl}

In this paper, we presented a machine learning framework to infer galaxy kinematic parameters from spatially unresolved spectra. The method employs a fully connected neural network trained on quantities derived from spatially resolved observations in \texttt{MaNGA}. With this approach, we are able to reliably separate galaxies with relatively high spin from those with lower values, providing a discriminator that may help identify galaxies likely to host supermassive black hole binaries.

This framework is designed to address two fundamental challenges in measuring galaxy kinematics. First, as galaxies are observed at increasing distances, they become unresolved point sources, requiring the inference of intrinsically multi-dimensional kinematic information from integrated spectra. We address this by training the network to learn how such information is encoded in spatially unresolved line profiles. Second, as demonstrated by fitting analytic profiles to the H$\alpha$ line, the finite spectral resolution limits the ability to extract kinematic information from traditional measures such as line widths alone. By contrast, the neural network leverages more subtle structure in the line profile, enabling it to extract information beyond what is accessible through parametric fits.

A systematic that we have not considered here is aperture bias~\citep{Kochanek2000}, whereby fixed-size fibers sample different physical regions of galaxies as a function of redshift, potentially modifying the observed line profiles and their relation to global kinematics. In the \citet{Bardati2024b} study, the discriminating quantity for hosts of supermassive black-hole binaries was found to be the spin parameter $\lambda_{R_e}$ evaluated at the half-light radius of the galaxy $R_e$. The \texttt{MaNGA} IFU set up, which we used to train our model, covers between 1.5 and 2.5 half-light radii of the observed galaxies~\citep{mangaApature}. To explore potential biases of applying this model to the \texttt{eBOSS} survey, we note that the field-of-view of the \texttt{eBOSS} optical fibers is 2 arcsec~\citep{eBOSS}. Using the effective radius of the Milky Way (5 kpc) and the fact that the angular diameter distance in standard $\Lambda_{\text{CDM}}$ cosmology peaks at $\sim1.8$~Gpc near $z\simeq 1.6$, a 2 arcsec fiber subtends a physical radius that rises from $\sim1.1~R_e$ at $z=0.4$ to $\sim1.5~R_e$ at $z=0.7$ and flattens near $1.7~R_e$ thereafter. The aperture sampled by an \texttt{eBOSS} fiber, therefore, falls within the range of physical apertures spanned by the \texttt{MaNGA} training sample and it varies only weakly with redshift across the range of interest. The magnitude of the aperture bias in this case is illustrated in Figure~\ref{fig:lamR}, where we plot the spin parameter as a function of aperture in terms of the half-light radius. This figure shows that using the \texttt{eBOSS} survey will introduce a bias in the inferred spin parameter of $\sim 20\%$, with an uncertainty of $\lambda_{R_e} \approx \pm20\%$. This aperture bias, however, is not large enough to prohibit  distinguishing slow- from fast-spinning galaxies according to the criteria in the study of \citet{Bardati2024b}.

Looking ahead, applying this approach to large spectroscopic surveys such as \texttt{DESI} could enable the estimation of $\lambda_R$ for millions of galaxies at higher redshift~\citep{DESI2026A}. This would allow statistical studies of galaxy kinematics well beyond the reach of integral field spectroscopy and may provide a new avenue for identifying merger remnants, including potential hosts of supermassive black hole binaries.

\section*{Acknowledgments}
P.W. gratefully acknowledges support from a graduate fellowship from S. Blackstock.  

Funding for the Sloan Digital Sky  Survey IV has been provided by the  Alfred P. Sloan Foundation, the U.S.  Department of Energy Office of  Science, and the Participating  Institutions. 

SDSS-IV acknowledges support and  resources from the Center for High  Performance Computing  at the  University of Utah. The SDSS  website is www.sdss4.org.

SDSS-IV is managed by the  Astrophysical Research Consortium  for the Participating Institutions of the SDSS Collaboration including  the Brazilian Participation Group,  the Carnegie Institution for Science,  Carnegie Mellon University, Center for  Astrophysics - Harvard \& 
Smithsonian, the Chilean Participation 
Group, the French Participation Group, 
Instituto de Astrof\'isica de 
Canarias, The Johns Hopkins 
University, Kavli Institute for the 
Physics and Mathematics of the 
Universe (IPMU) / University of 
Tokyo, the Korean Participation Group, 
Lawrence Berkeley National Laboratory, 
Leibniz Institut f\"ur Astrophysik 
Potsdam (AIP),  Max-Planck-Institut 
f\"ur Astronomie (MPIA Heidelberg), 
Max-Planck-Institut f\"ur 
Astrophysik (MPA Garching), 
Max-Planck-Institut f\"ur 
Extraterrestrische Physik (MPE), 
National Astronomical Observatories of 
China, New Mexico State University, 
New York University, University of 
Notre Dame, Observat\'ario 
Nacional / MCTI, The Ohio State 
University, Pennsylvania State 
University, Shanghai 
Astronomical Observatory, United 
Kingdom Participation Group, 
Universidad Nacional Aut\'onoma 
de M\'exico, University of Arizona, 
University of Colorado Boulder, 
University of Oxford, University of 
Portsmouth, University of Utah, 
University of Virginia, University 
of Washington, University of 
Wisconsin, Vanderbilt University, 
and Yale University.

\bibliography{export-bibtex}{}
\bibliographystyle{aasjournalv7}



\end{document}